\documentclass[5p,times,twocolumn]{elsarticle}

\usepackage{lineno,hyperref}
\usepackage{multirow}
\usepackage{float}
\usepackage{natbib}
\usepackage[ulem=normalem]{changes}
\usepackage{rotating}

\journal{Elsevier}

\let\linenumbers\nolinenumbers\nolinenumbers

\begin{document}

\begin{frontmatter}

\title{HeCTOr: the $^3$He Cryogenic Target of Orsay\\
for direct nuclear reactions with radioactive beams}

\fntext[fn1]{\emph{Present address:} Dipartimento di Fisica e Astronomia, Universit\`a di Padova, and INFN Laboratori Nazionali di Legnaro, Italy}

\author[IJCLAB]{F. Galtarossa\fnref{fn1}}
\cortext[mycorrespondingauthor]{Corresponding author}
\ead{franco.galtarossa@lnl.infn.it}

\author[IJCLAB]{M.~Pierens}
\author[IJCLAB]{M.~Assi\'e}
\author[IJCLAB]{V.~Delpech}
\author[IJCLAB]{F.~Galet}
\author[IJCLAB]{H.~Saugnac}
\author[Padova]{D.~Brugnara}
\author[GANIL]{D.~Ramos}
\author[IJCLAB]{D.~Beaumel}
\author[IJCLAB]{P.~Blache}
\author[IJCLAB]{M.~Chabot}
\author[IJCLAB]{F.~Chatelet}
\author[GANIL]{E.~Cl\'ement}
%\author[IJCLAB]{C.~Comault}
\author[LPC]{F.~Flavigny}
\author[GANIL]{A.~Giret}
\author[LNL]{A.~Gottardo}
\author[GANIL]{J.~Goupil}
\author[GANIL]{A.~Lemasson}
\author[LPC]{A.~Matta}
\author[GANIL]{L.~M\'enager}
\author[IJCLAB]{E.~Rindel}

\address[IJCLAB]{Universit\'e Paris-Saclay, CNRS/IN2P3, IJCLab, 91405 Orsay, France}
\address[Padova]{Dipartimento di Fisica e Astronomia, Universit\`a di Padova, and INFN-Padova, I-35131 Padova, Italy}
\address[GANIL]{Grand Acc\'el\'erateur National d'Ions Lourds (GANIL), CEA/DRF-CNRS/IN2P3, Bvd. Henri Becquerel, 14076 Caen, France}
\address[LPC]{Laboratoire de Physique corpusculaire de Caen, ENSICAEN CNRS/IN2P3, 14050 Caen, France}
\address[LNL]{INFN, Laboratori Nazionali di Legnaro, I-35020 Legnaro (Padova), Italy}

\begin{abstract}

Direct nuclear reactions with radioactive ion beams represent an extremely
powerful tool to extend the study of fundamental nuclear properties far
from stability.
These measurements require pure and dense targets to cope with the low beam
intensities.
The $^3$He cryogenic target HeCTOr has been designed to perform direct
nuclear reactions in inverse kinematics.
The high density of $^3$He scattering centers, of the order of 10$^{20}$
\,atoms/cm$^2$, makes it particularly suited for experiments where
low-intensity radioactive beams are involved.
The target was employed in a first in-beam experiment, where it was coupled
to state-of-the-art gamma-ray and particle detectors.
It showed excellent stability in gas temperature and density over time.
Relevant experimental quantities, such as total target thickness, energy
resolution and gamma-ray absorption, were determined through dedicated
Geant4 simulations and found to be in good agreement with experimental
data.

\end{abstract}


\end{frontmatter}

\linenumbers

\section{Introduction}

Direct nuclear reactions, where one or few nucleons are exchanged between
projectile and target, have proved to be extremely powerful tools to study
the structure of atomic nuclei~\cite{satchler} and to investigate a large
variety of astrophysical scenarios~\cite{bardayan}.
Their selectivity to the nature of the populated states and their sensitivity
to the transferred angular momentum help to obtain a detailed spectroscopy
of nuclei and access fundamental properties such as the quantum numbers
and single-particle character of ground and excited states.
These are the fundamental ingredients for the determination of nuclear
shell properties and their evolution across the nuclide chart.
The selectivity of the reaction mechanism results, in turn, in small cross
sections, typically of few mbarn.

With the development of radioactive ion beam (RIB) facilities (see
\cite{blumenfeld} and references therein) new possibilities are offered for
the study of the structure of exotic nuclei via direct reactions in inverse
kinematics, with the heavy unstable projectile impinging on the light target
at energies ranging from few to several hundreds MeV/u~\cite{hansen}.
RIB intensities are several orders of magnitude lower than stable beams,
and thus require high detection efficiencies and the use of thick targets
(typically few mg/cm$^2$).

For reactions involving hydrogen and deuterium, films of polypropylene like
CH$_2$ and CD$_2$ are commonly used, thanks to their simple handling.
The main drawback is the background generated by reactions on carbon, whose
contribution needs in most cases to be estimated with a dedicated measurement
at the expense of the effective beam time.
Morover the presence of heavier elements increases, along with the target
thickness, the energy and angular stragglings.

Part of these limitations can be overcome by developing pure targets where the
gas is confined in a small volume and cooled down to cryogenic temperatures.
In this way the density of scattering centers can be increased up to
a factor 50.
Many examples exist in the literature of hydrogen (H$_2$) and deuterium (D$_2$)
cryogenic targets designed for low- and intermediate-energy nuclear physics
experiments (see \cite{obertelli} and references therein).

Cryogenic targets of $^4$He~\cite{vandenbrand,ryuto,kendellen,hilcker} and
$^3$He~\cite{milner,kato,deschepper} are generally less widespread
than H$_2$ targets and the existing ones are not specifically designed to be
employed in experiments with low-energy ($\sim$\,10 MeV/u) and low-intensity
($\sim$\,10$^4$-10$^5$ pps) beams for transfer reactions.
In particular some are conceived for measurements, like for instance
electron scattering or photo-absorption, where temperatures lower than 4\,K
and/or target thicknesses of few cm are needed to reach areal densities of
tens or hundreds of mg/cm$^2$~\cite{vandenbrand,ryuto,kendellen,hilcker,kato}.
Others involve measurements with polarized $^3$He nuclei that can be performed
also in storage rings, where densities of scattering centers of
10$^{16}$-10$^{18}$\,atoms/cm$^2$ are sufficient~\cite{milner,deschepper}.

Valid alternatives conceived for low-energy nuclear reactions have been
recently developed.
Solid targets obtained by implantation~
\cite{geist,weissman,fernandez,godinho} have the advantage to be very compact
and easy to handle, whereas the employ of windowless and gas jet targets
~\cite{chymene,jensa,hippo} allows to strongly reduce background and
stragglings generated in the target windows. 
In both cases, though, target thicknesses of the order of 1\,mg/cm$^2$ are
hardly reached.

In this paper we report on a $^3$He cryogenic target where densities
of scattering centers of the order of 10$^{20}$\,atoms/cm$^2$, normally
sufficient to cope with low RIB intensities, are obtained with a target
thickness of few mm and an operation temperature higher than 4\,K.
These characteristics make it particularly suited to perform direct transfer
reactions, such as ($^3$He,d), ($^3$He,p), ($^3$He,n) and ($^3$He,$\alpha$),
with radioactive beams in inverse kinematics at bombarding energies ranging
from few to tens of MeV/u.
These reactions allow to tackle several interesting physics cases, among
which we mention shell evolution along neutron shell closures, neutron-proton
pairing in unstable nuclei, or proton drip-line physics.
The outreach of physics programs dedicated to such studies will strongly profit
from the installation of the target in fragmentation or next-generation
European post-accelerated ISOL facilities, SPES~\cite{spes} and
HIE-ISOLDE~\cite{isolde}, or the low-energy branch of FAIR~\cite{fair}.

\section{Constraints on the target design}\label{sec:requirements}

Properties of the reactions of interest, such as beam intensity and energy,
type of reaction and detected reaction products, strongly constrain the
design of the target and lead to several specific requirements.
In this Section we will report general considerations that should be taken
into account in the design of a cryogenic target for our purposes, in
Section ~\ref{sec:hector_design} we will describe in detail the specific
design of HeCTOr.

\paragraph{Dimensions}
The target has to be coupled to existing detectors which are usually arranged
in compact configurations around a traditional foil target of small dimensions.
For this reason the dimensions of the target and the associated cryogenic
equipment have to be as reduced as possible.
At the same time, the target radius must be large enough to give the
possibility to deal with both ISOL and fragmentation beams.
Typical ISOL beams have dimensions on target $\sigma$\,$\sim$\,2-3\,mm,
while fragmentation beams can feature beam spots at least a factor of 2 larger.

\paragraph{Thickness}
To measure transfer cross sections usually of the order of few mb with beams
of low intensity ($<10^5$\,pps), the choice of the target thickness must
balance the need of sufficient luminosity and good energy resolution.
These two requirements go in opposite directions.
High luminosity calls for the use of thick targets, of the order of few
mg/cm$^2$, which can be obtained with a large target cell and operating at
cryogenic temperatures.
Good energy resolution, which mainly depends on energy losses, stragglings
and re-interactions in the target, requires instead thin targets, usually of
the order of 100\,$\mu$g/cm$^2$ or less.
The best choice usually depends on the specific physics case.

\paragraph{Window thickness and material}
The target windows should be as thin as possible to reduce stragglings,
energy losses and contributions coming from background reactions.
At the same time they should be thick and elastic enough to withstand
considerable deformations due to the high difference in pressure between gas
inside the target cell and vacuum around it. 

\paragraph{Transparency}
The frame surrounding the gas target must have an opening around the target
as large as possible to allow the emitted particles to reach the detectors,
thus increasing the detection efficiency.
Very large openings would require thicker windows and are difficult to obtain
due to mechanical constraints.
The target can then be designed to have a larger opening in the hemisphere
(forward or backward) where the laboratory cross section of the reaction
channel of interest is higher.
The transparency to electromagnetic radiation (X and $\gamma$ rays) emitted
by the reaction products can be another important requirement.
This once again calls for a reduction and an accurate choice of the materials,
surrounding the gas target, that might ``shadow'' the $\gamma$-ray detector,
absorbing part of the electromagnetic radiation emitted by the reacting
nuclei.

\paragraph{Target versatility}
A good versatility is an essential property to cover a large number of
physics cases without deeply modifying the set-up between different
experiments.
For instance, the possibility to change the gas inside the target cell, to
modify its pressure and temperature and to switch forward and backward sides
gives considerable flexibility to the target and allows to perform a large
variety of reactions.

\section{HeCTOr design and operation}\label{sec:hector_design}

In this Section we will describe the characteristics and operation principles
of HeCTOr, the $^3$He cryogenic target developed at IJCLab and designed
to perform direct reactions with radioactive beams in inverse kinematics
from few to tens of MeV/u.
All the requirements described in Section~\ref{sec:requirements} have been
carefully taken into account in its design.
In Section~\ref{sec:experiment} we will present the performance during the
first in-beam experiment, where the target was coupled to the MUGAST
~\cite{mugast_campaign} silicon array, the AGATA~\cite{agata,clement}
$\gamma$-ray array and the VAMOS~\cite{vamos} large-acceptance magnetic
spectrometer.

\paragraph{The target cell}

\begin{figure}[ht]
\begin{center}
\includegraphics[width=\columnwidth]{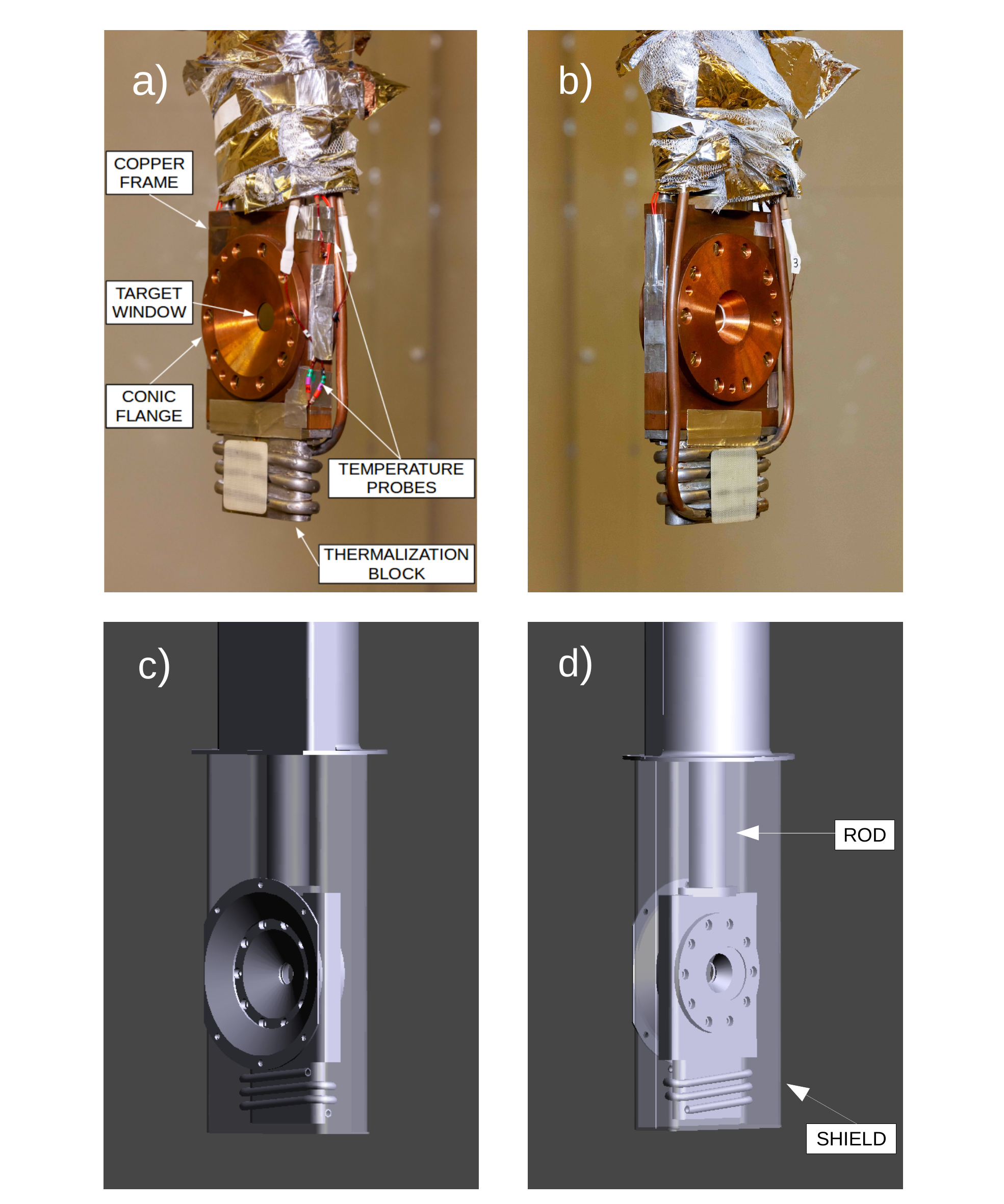}
\end{center}
\caption{(Top) Photographs of the target cell without the thermal shield:
a) backward side, from which the beam enters the target; b) forward side.
(Bottom) CAD drawings of the target cell where the shield is added as a
semi-transparent layer: c) backward side; d) forward side.
The rod connecting the target cell to the LHe tank is also visible.}
\label{fig:target_cell}
\end{figure}

\begin{figure}[ht]
\begin{center}
\includegraphics[width=\columnwidth]{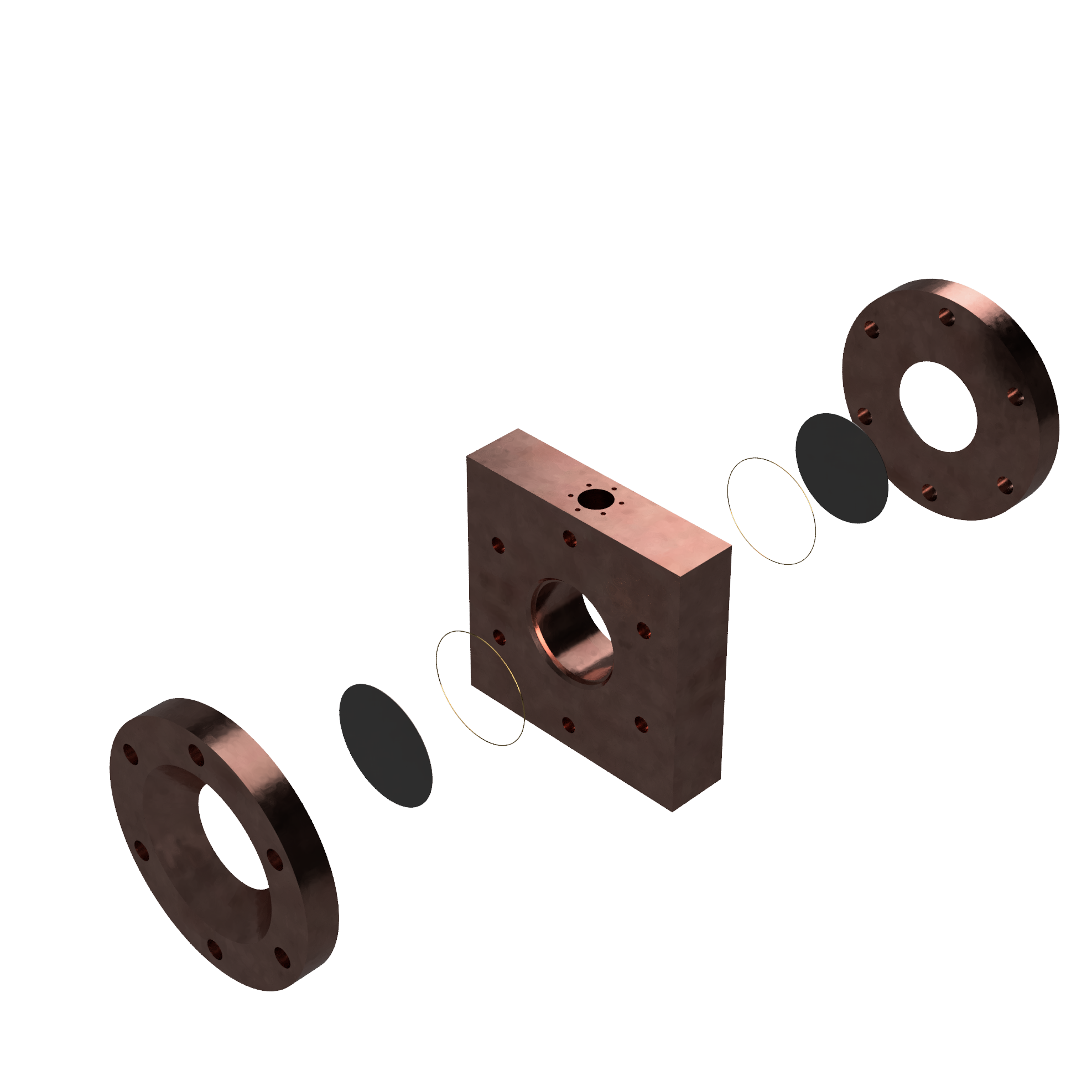}
\end{center}
\caption{Magnified view of the target cell. From left to right: conic
flange, Havar window, indium wire and target frame.
In the forward side the sequence is repeated in inverted order.}
\label{fig:magnified_view}
\end{figure}

The target cell, shown in Fig.~\ref{fig:target_cell}, is composed of a main
copper frame to which two conic copper flanges are screwed, one on each
side of the frame.
A thin foil, serving as target window, is then glued to each flange with
epoxy resin glue.
The diameter of the smaller base of the cone is 16\,mm, the frame thickness
is 15\,mm and the nominal target thickness is 3\,mm.
The nominal target thickness is defined as the distance between the two
windows when $^3$He gas is at the nominal operation pressure of
$\sim$\,1\,bar inside the target, neglecting window deformations.
An indium wire is placed between the conic flange and the frame to avoid
leakages at cryogenic temperatures between the $^3$He circuit and the
vacuum around it.
The different parts composing the target cell can be seen in the magnified
view of Fig.~\ref{fig:magnified_view}.

A coil-shaped copper thermalization block is brazed at the bottom of the
frame to ensure good thermal contact and operates as a heat exchanger to
cool down the target by means of liquid $^4$He flowing through it.
The block contains a pipe with an inner and outer diameter of 4\,mm and
6\,mm, respectively.

The conic flange has a larger angular opening (130$^{\circ}$) in the backward
hemisphere to adapt for stripping reactions\footnote{The expressions
\emph{stripping} and \emph{pick-up} may be confused when the reaction occurs
in inverse kinematics.
Here and later in the text we always refer these expressions to the light
particle, so in a stripping (pick-up) reaction the light target
particle loses (gains) nucleons.}, like ($^3$He,d) or ($^3$He,p),
where the light recoiling particles are mainly scattered at
$\theta_{\mathrm{lab}}$\,$>$\,90$^{\circ}$.
The target could also be filled with $^4$He when necessary.
For pick-up reactions, where the ejectiles are mainly scattered forward, like
($^3$He,$^4$He) but also ($^4$He,$^6$He) or ($^4$He,$^8$Be), it can be rotated
to adapt to the specific reaction. 

For the target windows the considerations outlined in Section~
\ref{sec:requirements} led to the choice of 3.8-$\mu$m thick foils of Havar,
an alloy of cobalt (42\,\%), chromium (20\,\%), nickel (2.7\,\%), tungsten
(2.2\,\%) and other materials in smaller percentage. 
The corresponding areal density of each foil is 3.15\,mg/cm$^2$.
The typical pressure of gas in the target is 1\,bar absolute.
The relatively high pressure and the elasticity of Havar induce a
deformation of the windows, defined as the maximum distance along the beam
direction between the nominal target thickness and the Havar window.
Such deformation is measured using a micrometric probe placed in the center
of the target window.
A first measurement is made with the same (atmospheric) pressure in the gas
target and around it.
A second one is carried out setting a pressure of 2 bars inside the target
and atmospheric pressure around it, in order to have a pressure difference of
1 bar, as in the case of nominal operation.
The difference between the two values gives the window deformation,
measured to be (0.70\,$\pm$\,0.05)\,mm at room temperature.
Thanks to the properties of Havar, a similar value is expected also at
cryogenic temperatures (5\,-\,6\,K).
When it is cooled down to such temperatures, the equivalent areal density
of $^3$He gas is $\sim$\,2\,mg/cm$^2$ and the density of scattering centers
is $\sim$\,3.5\,$\cdot\,10^{20}$\,atoms/cm$^2$,
about 2-3 orders of magnitude higher than what can be obtained with typical
solid $^3$He-implanted or gas jet targets.
Since a small gradient of temperature is present along the cooling circuit
in the gas cell region, the target temperature is determined as the average
of the temperatures measured by two probes, visible in Fig.~
\ref{fig:target_cell} (top), placed at the top and bottom of the target cell.

\paragraph{Cryostat description}

\begin{figure}[ht]
\begin{center}
\includegraphics[width=\columnwidth]{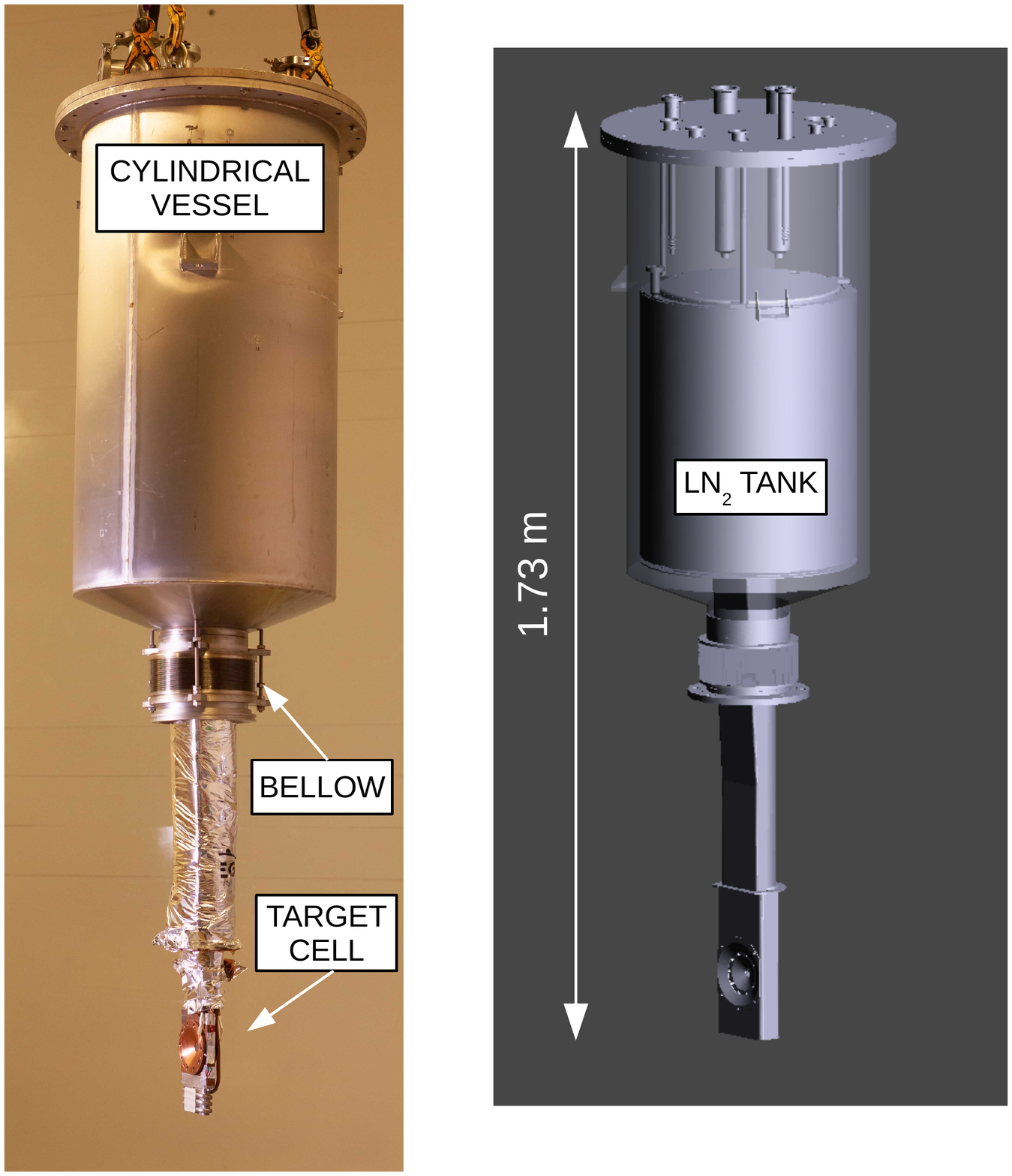}
\end{center}
\caption{(Left) Photograph of HeCTOr.
(Right) CAD drawing of HeCTOr.
The cylindrical vessel is semi-transparent to allow the visualization
of the LN$_2$ tank.
The two asymmetric hemicylindrical sheets enclosing the glass-epoxy
rod connect the bellow to the target cell.}
\label{fig:cryostat}
\end{figure}

\begin{figure*}[ht]
\begin{center}
\includegraphics[width=\textwidth]{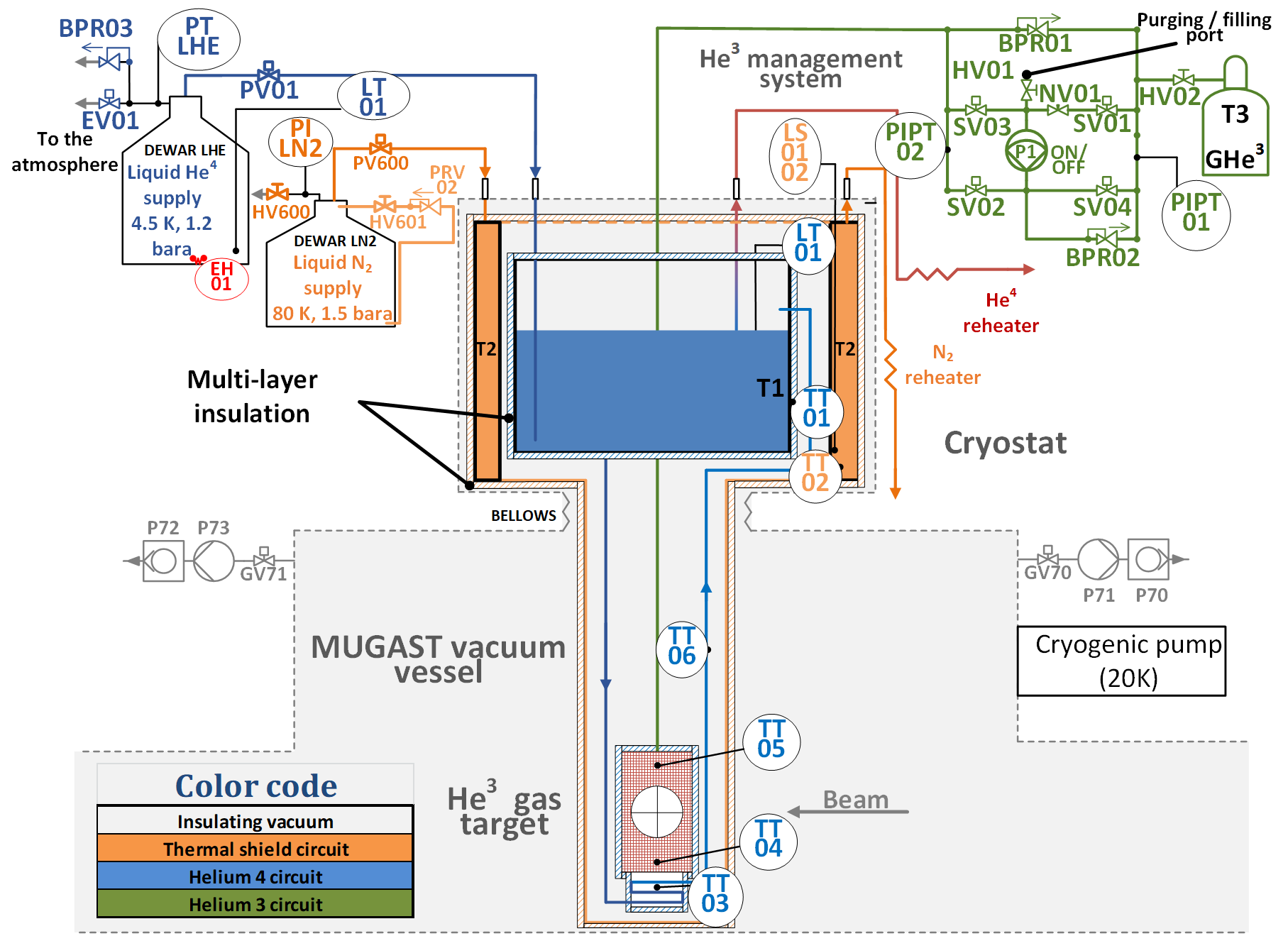}
\end{center}
\caption{Scheme illustrating the piping and instrumentation diagram of
HeCTOr (see text for details).}
\label{fig:scheme}
\end{figure*}

The small volume of $^3$He gas contained in the target cell is cooled down
to cryogenic temperatures by circulating liquid $^4$He (LHe) around the
target region.
The LHe is initially injected in the system from the top of the cryostat,
defined as the assembly of HeCTOR, visible in Fig.~\ref{fig:cryostat},
and the MUGAST vacuum chamber (refer also to Fig.~\ref{fig:scheme}, where
the cryostat is enclosed by the dashed gray line).
The LHe is then stored in the tank, or phase separator (T1), from which
it can finally reach the target.
To minimize the consumption and allow the LHe to maintain the cryogenic
temperature, a thermal shield is provided by liquid nitrogen (LN$_2$) at
atmospheric pressure and at a temperature of about 77\,K, measured with
a PT100 (TT02) placed at the bottom of the LN$_2$ tank (T2).

The target is insulated from the external environment through
the outer wall of the cryostat, composed in the upper part of a
cylindrical vessel made of AISI 304L stainless steel, with an outer
diameter of 450\,mm and a length of about 900\,mm, and in the lower part
of the MUGAST vacuum chamber.
The LN$_2$ and LHe circuits, comprising the LN$_2$ and LHe tanks and the
$^3$He gas target, are hosted inside the outer wall.
The target is supported by a glass-epoxy rod, a heat-insulating material,
with a length of 812\,mm, an inner diameter of 28\,mm and an outer diameter
of 30\,mm, attached to the LHe tank T1.
The total height of HeCTOr, from the top of the vessel to the bottom
of the target cell, is approximately 1.73\,m.

The operation scheme of the cryostat is shown in Fig.~\ref{fig:scheme}.
In this figure the outer wall, which can be seen as the first layer of
insulation, is represented by the dotted gray line.
A second layer, in orange, is maintained at LN$_2$ temperature and acts as
a thermal shield to insulate the cold gas target from radiations from
the vacuum vessel at room temperature.
This shield is composed at the top of the LN$_2$ tank and its cover
(orange dashed line) and in the central part of two 1.5-mm thick sheets
of copper with an asymmetric hemicylindrical-like shape that connect the
tank T2 to the target cell and enclose the glass-epoxy rod (see
Fig.~\ref{fig:cryostat} right).
These sheets are mechanically and thermally anchored to the tank and
cooled by thermal conduction.
The bottom of the thermal shield covers the target cell, as shown in
Fig.~\ref{fig:target_cell} (bottom), to protect the cold mass from the
room temperature.
A cut of the sheets is made at the level of the target windows to ensure
transparency to the beam. 

The thermal shield is covered by a multi-layer insulation (MLI), or
super-insulation), where each layer acts as a thermal shield, reducing
the radiated heat transfer from the vacuum vessel to the shield.
MLI consists of stacks of double-aluminized Mylar to reflect heat radiation,
separated by low-conduction spacers to limit heat transfer by conduction.
The temperature of each layer is left floating and then depends on the
global thermal equilibrium of the system.
MLI reduces the heat radiated to the cold mass by a factor $\sim$\,200 with
respect to the case where it is not present.
The upper part of the shield and the cold mass are covered by 30 and
10 layers of MLI, respectively.

The cooling circuit includes the tank T2 with an annular shape and a
capacity of about 13\,L, a filling and an exhaust line.
The tank T2 is supplied with LN$_2$ flowing out of a Dewar connected via a
flexible vacuum-insulated transfer line on top of the cryostat.
The cold vapors are warmed at the outlet of the circuit before returning to
the atmosphere, avoiding ice formation at the exhaust of the cryostat.

The cold mass, or Helium-4 circuit, shown in blue in Fig.~\ref{fig:scheme}
and indicating the $^3$He gas target and its active cooling circuit, is
located at the center of the system.
LHe at atmospheric pressure is used to maintain the target temperature
below 7\,K.
The circuit includes the cylindrical phase separator T1 with a useful
capacity of about 11\,L, filled by a LHe Dewar via a vacuum-insulated
and flexible transfer line.
The cold vapors return to the atmosphere through the outlet pipe where the
presence of an electrical heater avoids the formation of ice.
The $^3$He cryogenic gas target is cooled down by means of LHe coming
from T1 and passing through the coil-shaped thermalization block brazed
at the bottom of the copper frame.
Cold vapors from the boiling helium are returned to the top of the phase
separator and flow out of the cryostat via the tank outgassing line.

The Helium-3 circuit includes, outside the cryostat, a set-up based on a
dry primary vacuum pump to inject or recover the gas in the target inside
the cryostat and a tank with a volume of 90\,L for the $^3$He gas storage
at sub-atmospheric pressure (28\,mbar).
To fill the target with $^3$He gas, the pump is turned on and the valves
SV01 and SV02 are opened.
The needle valve NV01 moderates the flow to the target to limit sudden
pressure changes.
For the gas recovering procedure the pump is switched on and the valves
SV03 and SV04 are opened.
The back-pressure regulator BPR01 prevents the target from overpressure,
while BPR02 protects the pump.
The $^3$He circuit in the cryostat acts as a filling and recovery line to
set the pressure in the target.

The isolation vacuum level is measured by means of a combined Pirani and
Penning vacuum gauges.
We have available a test bench with a dedicated small chamber where the
target can be accommodated.
In this configuration, the vessel with a volume of about 156\,L is
pumped by a turbomolecular pump with a pumping speed of 150\,L/s in series
with a primary rotary vane pump with a pumping speed of 8\,m$^3$/h,
allowing to reach a vacuum of about 10$^{-5}$\,mbar absolute at 300\,K.
When the system is cold, there is a gain of an order of magnitude on the
vacuum level.
The LN$_2$ Dewar is provided with a self-pressurization system and the
transfer is controlled by an ON-OFF pneumatic valve (PV600).
A high and low thresholds are set on the LN$_2$ level in the tank to
automatically stop and start the transfer procedure.
A PT100 thermometer (TT02) is installed at the bottom of the tank to check
in particular the first filling during the cool-down process, since it
represents the most delicate step.
The LHe Dewar is equipped with a heater (EH01), a back-pressure regulator
(BPR03) and a solenoid valve (EV01) to pressurize the Dewar.
When a transfer is needed EV01 is closed, forcing the gas to flow through
the back-pressure regulator BPR03 with a pressure inside the Dewar set to
about 1.1\,bar, and the heater is activated to produce vapors.
This increases the pressure in the Dewar, allowing a transfer of LHe via
the activation of the pneumatic valve PV01.
A pressure transmitter (PTLHE) monitors the pressure of the Dewar. 
A LHe level sensor measures the quantity of LHe remaining in the Dewar
while a second one monitors the level of the liquid in the tank inside the
cryostat to start and stop the transfer.
Cernox$^{\mathrm{TM}}$ 1050 AA thermometers are located at the bottom of
the LHe tank (TT01), on top and at the bottom of the target (TT04 and TT05),
on the thermalization block (TT03) and on the outgassing of the cooling
circuit of the target (TT06).
The pressure of the $^3$He storage tank and of the target are measured by 
piezoresistive pressure sensors (PIPT01 and PIPT02, respectively).

The acquisition and actuators controls are performed by means of a
multiplexer Agilent 34970A equipped with two 34901A cards to measure
resistances in four-wires mode, LHe level (4-20\,mA) and pressure sensors
(0-10\,VDC).
A 34903A relays card is used to activate the heater of the LHe Dewar and
the ON-OFF valves.
A LabVIEW program communicates with the multiplexer to manage LN$_2$ and
LHe transfers and provides system supervision and data logging.

\paragraph{Cryostat operation}

The first step in the procedure to cool down the target is to pump the
vacuum vessel below 10$^{-5}$\,mbar at room temperature.
In the experimental configuration described in Sec.~\ref{sec:experiment}
the target is accommodated in the MUGAST reaction chamber, with a volume
of 515\,L.
To make the vacuum we employed a primary pump Agilent IDP-10, two
turbomolecular pumps Agilent Turbo-V750 TwisTorr (model 9696018) with a
pumping speed of 700\,L/s for N$_2$ and a cryogenic pump, composed of a
compressor unity Coolpak 4000 Leybold and a cold head Cryo-plex 8 model
350 Oxford instruments.
This pumping system allowed to reach a vacuum level of 10$^{-6}$\,mbar.

The LN$_2$ circuit does not need a specific conditioning.
By opening the PV600 valve, LN$_2$ starts to flow out of the Dewar
towards the cryostat and the air remained in the circuit is flushed.
For the LHe circuit, the high risk of blocking due to the possible
solidification of air in the circuit prevents from carrying out the same
simple procedure.
The circuit is then conditioned by pumping it below 1\,mbar and injecting
$^4$He gas slightly above the atmospheric pressure.
This operation is then repeated at least three times.
The $^3$He circuit is pumped down to 10$^{-6}$\,mbar via the purging port
(HV01) with all solenoid valves opened and isolated by closing HV01 and
then the solenoid valves.

The thermal shield is cooled down to 77\,K by means of LN$_2$.
When the nominal working mode is reached the temperature is maintained
stable keeping the LN$_2$ level between the high (LS01) and low (LS02) level
until the end of the experiment.
The $^4$He circuit is cooled down by pressurizing the Dewar and opening
the transfer valve PV01 until the tank T1 is filled.
The cooling circuit of the target starts and, when the target reaches an
equilibrium temperature of about 7\,K, the target operation can start.
The quantity of LHe needed to cool the target down to this temperature is
about 200\,L.

Once the cryostat is working in nominal conditions, the hand valve HV02
is opened and the target is filled with $^3$He gas to atmospheric
pressure and maintained during the experiment.
The power consumption of the target at equilibrium is $\sim$\,1.7\,W.

After the experiment, the LN$_2$ and LHe supply are stopped simultaneously
and the cryostat begins to warm up until the room temperature is reached.
The $^3$He gas is recovered and stored in the tank at low pressure.
The $^3$He gas is not purified after each operation but only when
a possible contamination with other gases might have occurred during the
target operation.
In such a case the tank can be connected to a specific purification tool,
composed of a cold trap, to eliminate impurities from the gas.

\section{Performance during the experiment}\label{sec:experiment}

HeCTOr has been employed for the first time in an in-beam experiment in
June 2019 at GANIL~\cite{ganil} in Caen, France.
The aim of the experiment was the determination of the proton occupancies
in the ground state of the N=28 $^{46}$Ar isotope, which can be achieved
measuring the differential cross sections of the deuterons emitted in the
proton stripping reaction $^{46}$Ar($^3$He,d)$^{47}$K.

In this Section we will describe the experimental set-up employed in GANIL
and the performance of HeCTOr.
This will allow us to extract and discuss important experimental quantities
that must be evaluated when designing and analyzing an experiment with such
a target, in particular stability over time (Section~\ref{sec:exp_cond}),
energy losses (Section~\ref{sec:ice_effect}), energy resolution
(Section~\ref{sec:energy_res}), and transparency to $\gamma$ rays
(Section~\ref{sec:agata}).

\subsection{Experimental conditions and beam monitoring}\label{sec:exp_cond}

The SPIRAL1~\cite{spiral} $^{46}$Ar beam, produced by the fragmentation of
$^{48}$Ca, is re-accelerated at 10\,MeV/u and impinges on the $^3$He gas
target with an intensity of approximately 4\,$\cdot$\,10$^4$\,pps.
Along the beam line, about 2\,m before the target, a beam tracking device,
CATS~\cite{CATS}, provides beam rate and profile monitoring and a
signal for time-of-flight measurement.

\begin{figure}[ht]
\begin{center}
\includegraphics[width=\columnwidth]{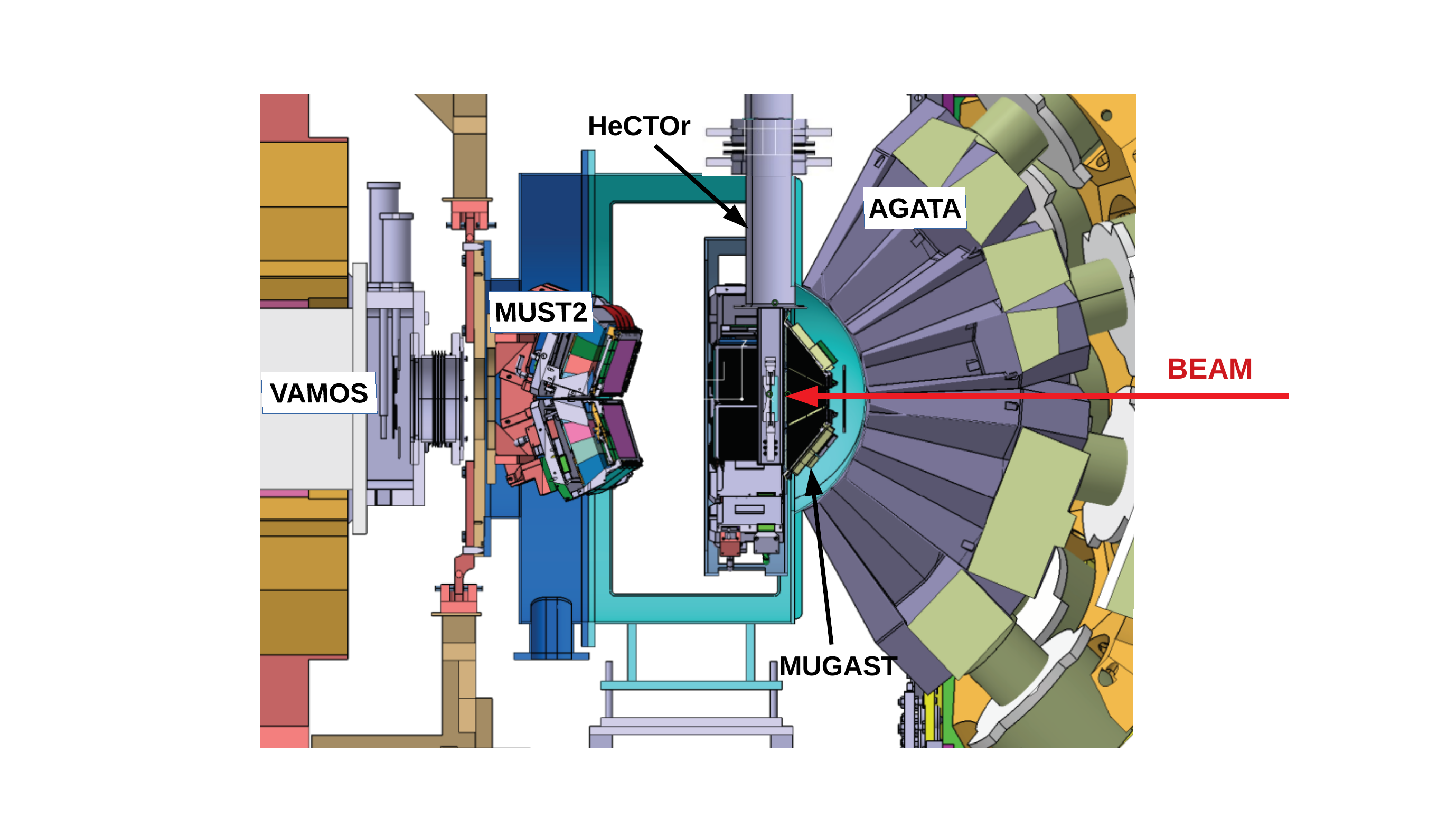}
\end{center}
\caption{A drawing of the experimental set-up at GANIL.}
\label{fig:GANIL_setup}
\end{figure}

The detection set-up, shown in Fig.~\ref{fig:GANIL_setup}, is composed of the 
segmented silicon array MUGAST~\cite{mugast_campaign}, coupled to the
high-resolution segmented HPGe array AGATA~\cite{agata,clement} and the
large-acceptance magnetic spectrometer VAMOS~\cite{vamos}.
This set-up allows the simultaneous detection in coincidence of both
reaction partners and the $\gamma$ rays emitted by the heavy recoil.
The MUGAST array is composed of 12 Double-sided Strips Silicon Detectors (DSSD)
surrounding the target: 5 trapezoidal and one annular detector in the backward
hemisphere, two square detectors close to 90$^{\circ}$ and 4 MUST2~\cite{must2}
detectors at forward angles.
The light ejectiles (deuterons in this case) are identified
through E-ToF correlation and their scattering angle and energy are measured
with an angular resolution better than 1\,degree and an intrinsic energy
resolution of $\sim$\,40\,keV.
The excitation energy of the heavy recoil ($^{47}$K) is then determined via
two-body kinematics.
AGATA is placed upstream with respect to MUGAST and detects the $\gamma$
rays emitted by the reacting nuclei.
The typical efficiency of AGATA at 18\,cm from the target, with MUGAST
installed, is 6-7\,\% for 1.3-MeV $\gamma$ rays.
VAMOS is placed at 0$^{\circ}$ for the detection of the heavy residue and
its identification in atomic number and mass.
The coincidence with VAMOS allows to strongly suppress contributions from other
reaction channels, in particular fusion-evaporation reactions on the target
windows.
Further details on the detection set-up can be found in Ref.
~\cite{mugast_campaign}.

The target system, inserted in the chamber from the top with a crane and secured
to a mechanical support attached to the chamber, is placed 25\,mm downstream
with respect to the nominal target position, due to mechanical constraints,
and cooled down to $\sim$\,6\,K before the beginning of the beam time.

The $^{46}$Ar beam ions passing through the target are detected in the focal
plane detectors of VAMOS and separated depending on their magnetic rigidity
B$\rho$, according to:

\begin{center}
\begin{equation} \label{eq:brho}
B\rho = 3.105\cdot\beta\gamma~\frac{A}{q}
\end{equation}
\end{center}

where A and q are the mass number and atomic charge of the ions, respectively,
$\beta = v/c$ (v is the velocity of the ions) and $\gamma=(1-\beta^2)^{-1/2}$.

\begin{figure}[ht]
\begin{center}
\includegraphics[width=\columnwidth]{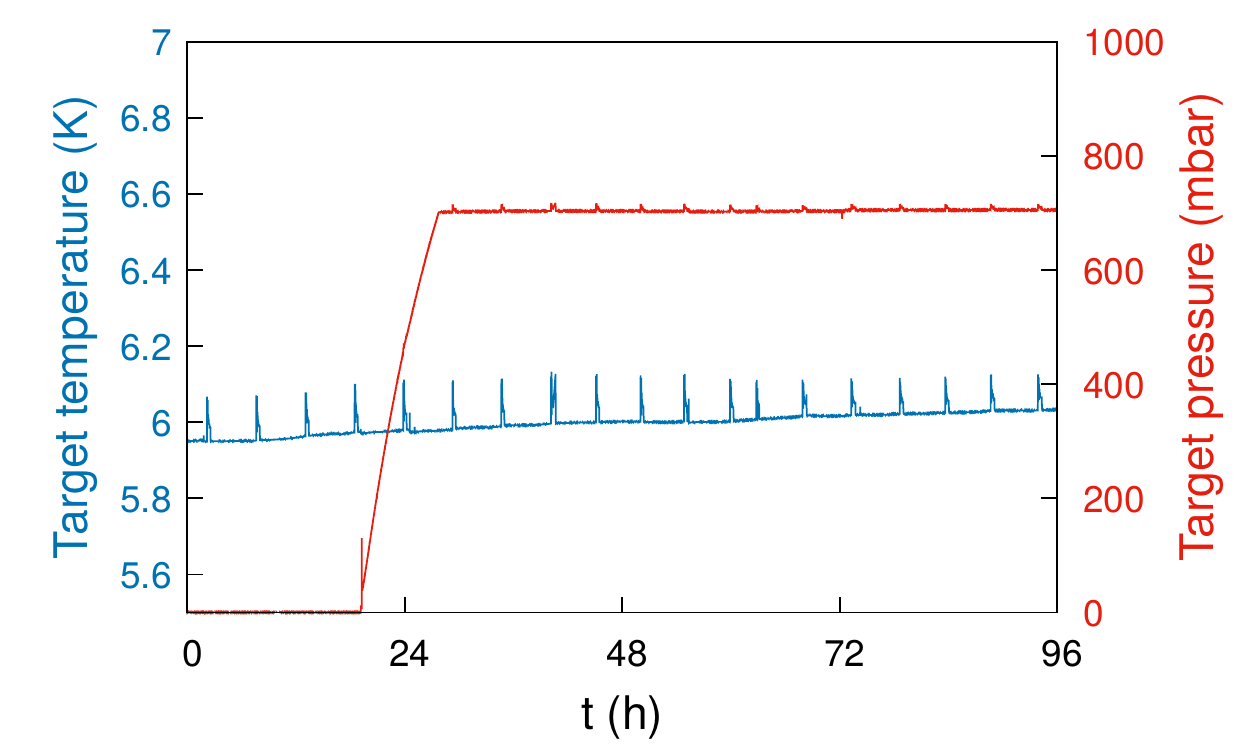}
\end{center}
\caption{Plot of the temperature in K (blue) and pressure in mbar (red) of
the target during 4 days of experiment.
The small spikes occurring approximately each 6 hours, and visible mainly
in the temperature curve, are in correspondence of the filling of the LHe
tank.}
\label{fig:plot_Tp}
\end{figure}

\begin{figure}[ht]
\begin{center}
\includegraphics[width=\columnwidth]{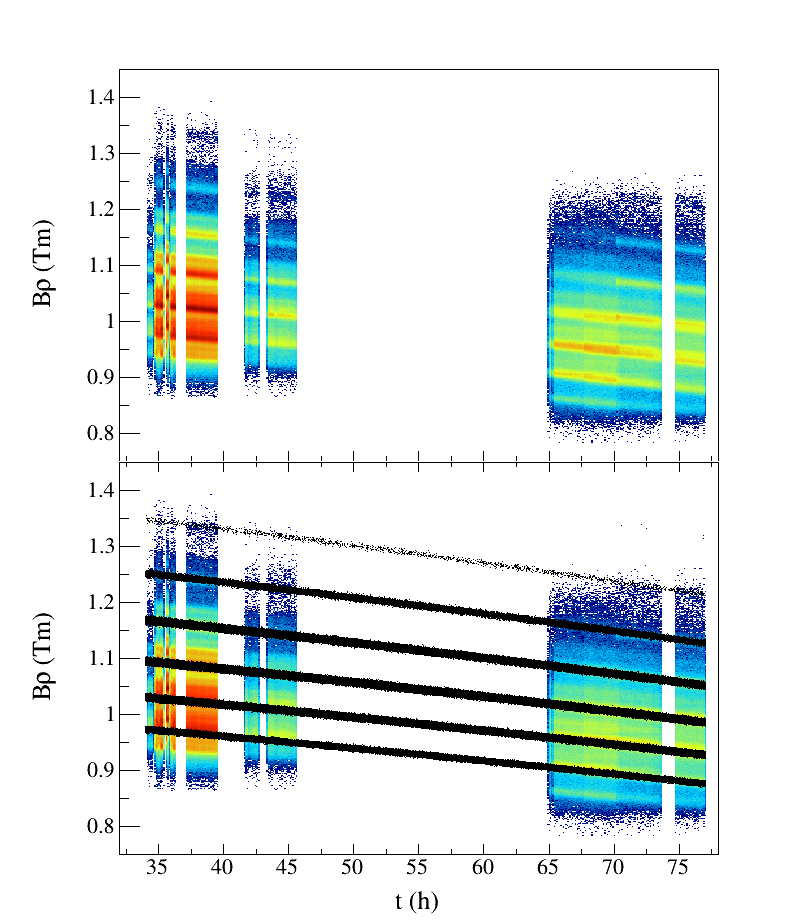}
\end{center}
\caption{(Top) Experimental B$\rho$ as a function of time for $\sim$\,40\,h
after the filling of the target with $^3$He.
(Bottom) Same as the top panel but with the simulated B$\rho$ distribution
superimposed in black.}
\label{fig:trend}
\end{figure}

When the target is empty the beam particles lose energy passing through the
CATS detector and the two Havar windows.
When the target is filled with $^3$He the energy loss of beam and light
ejectiles in the gas has to be considered as well.
By monitoring the temperature and pressure of the gas in the target cell
during the whole experiment, possible changes in the gas density can be
identified and accounted for.
Figure~\ref{fig:plot_Tp} shows the trend of the target temperature (in blue)
and pressure (in red) as a function of time during 4 days of experiment.
The t\,=\,0 reference has been arbitrarily set at the midnight of the day
when the experiment started and has been kept the same in Figure
~\ref{fig:trend}.

We started to fill the target with $^3$He at t\,$\sim$\,19\,h, reaching the
final pressure of 700\,mbar at t\,$\sim$\,27.5\,h.
During the filling and the rest of the experiment the target temperature
remained stable and close to 6\,K.
At this temperature and pressure the $^3$He density is 4.5\,mg/cm$^3$.
The maximum target deformation, measured to be 0.7\,mm per side at 1\,bar
pressure, scales down to 0.5\,mm for a pressure of 700\,mbar.
The corresponding areal density is $\sim$\,2\,mg/cm$^2$ and the density of
$^3$He scattering centers is $\sim$\,3.5\,$\cdot\,10^{20}$\,atoms/cm$^2$.
The level of vacuum in the chamber, stable during the whole experiment, was
1.5\,$\cdot$\,10$^{-6}$\,mbar.
The small increases in temperature and pressure occurring approximately each
6 hours, visible in Fig.~\ref{fig:plot_Tp}, are due to the change of pressure
occurring in the thermalization block when the LHe transfer process starts.
Such change temporarily interrupts the LHe circulation in the pipe, causing
a small increase of the temperature.
Then, when LHe starts to flow again, the temperature eventually decreases.

The target thickness can be determined and monitored over time during the
experiment by measuring the change in beam ion velocity after the target with
VAMOS, according to Eq.~\ref{eq:brho}, and computing the corresponding energy
loss in the target.
The top panel of Fig.~\ref{fig:trend} shows the experimental B$\rho$ for a time
interval of $\sim$\,40\,h after the target filling procedure has been completed.
Six main structures are visible, corresponding to the most intense charge states
of beam ions detected in VAMOS (from 13$^+$ to 18$^+$).
The intervals where no B$\rho$ information is available represent periods of time
where the beam was not delivered.
The B$\rho$ is clearly not constant over time but decreases smoothly, in turn
indicating a progressive decrease of the velocity of beamlike ions reaching the
VAMOS focal plane.
This behavior might be explained assuming that the target thickness, instead of
being constant, slightly increases over time.
Since the temperature and pressure in the target are stable (see Fig.
~\ref{fig:plot_Tp}), the increase of target thickness could be associated to the
thickening of layers of ice forming on the target windows through the deposit
of frozen gas, probably coming from the MUGAST cooling system.
The presence of other contaminants, like air coming from out of the reaction
chamber or small quantities of gas coming from the gas detectors of the VAMOS
focal plane, cannot be a priori excluded.

A rough estimation of the expected rate of ice growth can be obtained through
simple considerations of kinetic theory of gases, in the approximation of ideal
gas.
Under the assumption that the velocity of the gas particles in the surroundings
of the target follows the Maxwell's distribution, it can be shown that the number
of collisions per unit time and unit area is given by:

\begin{equation}\label{eq:Nc}
N_c = \frac{1}{4}n\bar{v} = \frac{P}{\sqrt{2\pi mk_BT}}
\end{equation}

where $n = P/(k_BT)$ is the volumic density of molecules, $\bar{v}\,=\,
\sqrt{8k_BT/(\pi m)}$ is the mean velocity of the Maxwell distribution, k$_B$
is the Boltzmann constant, T and P the gas temperature and pressure,
respectively, and m the mass of the gas molecules.
Due to the very low temperature of the surface, the molecules will stick to it
after the collision.
A monolayer of thickness d$_0$ is then formed in a time:

\begin{equation}\label{eq:monolayer}
t_m = \frac{1}{AN_c} = \frac{4}{d_0^2 n \bar{v}}
\end{equation}

where A\,=\,d$_0^2$ is the area of a molecule.
The volume of a molecule in the ice can be approximated as
$d_0^3 = M/(N_A\rho)$ where M is the molar mass, N$_A$ the Avogadro number
and $\rho\sim$\,10$^3$\,kg/m$^3$ the ice density.
The resulting monolayer thickness is d$_0\sim$\,3\,\AA.
With the appropriate substitutions, Eq.~\ref{eq:monolayer} can be rewritten as:

\begin{equation}\label{eq:rate_growth}
t_m = \frac{\sqrt{2\pi k_BT}}{P}\rho^{2/3}\left(\frac{N_A}{M}\right)^{1/6}
\end{equation}

For the present experiment P\,=\,1.5\,$\cdot\,10^{-6}$\,mbar, T\,=\,300~K.
Assuming in a first approximation that the gas is coming only from the cooling
system of MUGAST, its composition is 50\,\% water (M$_\mathrm{{H_2O}}$\,=\,18.02
\,g/mol) and 50\,\% ethyl alcohol (M$_\mathrm{{C_2H_6O}}$\,=\,46.07\,g/mol),
so an average value M\,$\sim$\,32 can be considered.
A detailed study of the gas composition would require a dedicated measurement
and is beyond the aim of this simple calculation.
Substituting these values in Eq.~\ref{eq:rate_growth}, we obtain
a monolayer formation time t$_m$\,$\sim$\,1.6\,s, corresponding to a rate of
thickness growth per window of $\sim$\,22\,$\mu$m/day.
This value is not expected to change dramatically if some impurities are
added to the gas.
Different inputs for M and $\rho$, to account for varied gas compositions,
result in values of thickness growth per day which do not deviate by more than
50\,\% from the present value.

The ice formation rate can be monitored experimentally by measuring the
B$\rho$ of beamlike ions in VAMOS and computing the correspondent energy loss
in the target.
To this aim, we performed simulations with the Zgoubi code~\cite{zgoubi},
where the incoming ions are propagated through the optics of the spectrometer
and the energy losses in the different materials along the ion path are taken
into account using SRIM~\cite{srim}.
In the simulation we can vary the ice thickness, under the assumption that the
ice forms at the same rate on the front and back window, until a good matching
between simulation and data is found.
The result is shown in the bottom panel of Fig.~\ref{fig:trend}, where the
simulated B$\rho$ for the different charge states is superimposed in black on
the experimental data.
The ice thickness at the beginning of this plot (t\,$\sim$\,35\,h) has been set
to 37\,$\mu$m to reproduce the B$\rho$ value after all the other energy losses
of the beam (in CATS, target windows and gas) have been considered as well.
The experimental trend is correctly reproduced by setting in the simulation a
rate of ice formation on the target windows of a total of 22\,$\mu$m/day
(11\,$\mu$m per window), in fair agreement with the simple calculation discussed
above.
The total ice thickness at t\,$\sim$\,75\,h is then approximately 74\,$\mu$m.
The simulation also shows that the B$\rho$ structures for t\,$>$\,65\,h are the
expected extension of those at t\,$<$\,46\,h, when the proper energy loss in the
target is considered.

\subsection{Impact of the ice formation}\label{sec:ice_effect}

The energy spectrum of the deuterons emitted in the present reaction extends
down to 2.5\,MeV at the most backward laboratory angles, which is also the
angular region where the transfer cross sections are higher.
This feature can be common to other stripping reactions at similar bombarding
energies, of course depending on the reaction Q value. 
These particles lose part of their energy passing through the gas composing the
target, the Havar window and the ice layers on the window and may eventually
reach the detector with not enough energy to overcome the detection threshold.
For this reason the increase of ice thickness on the target windows over time
turns out to be a highly undesired effect.

To give some numbers, 2.5-MeV deuterons lose about 1\,MeV of energy in
40\,$\mu$m of ice, the same amount of energy that they lose in about 9\,$\mu$m
of Havar or in 5.5\,mm of $^3$He at the temperature and pressure reached in the
present experiment (energy losses computed with LISE++~\cite{lise}).
These numbers also help to highlight the importance of the correct balance
between target thickness and energy losses for this kind of reactions.
Higher density of scattering centers ($\sim$ 30 times more) could be obtained,
for instance, operating the target with liquid $^3$He at $\sim$\,4\,K,
condition that would allow to perform experiments similar to the one presented
in this article with beam intensities of less than 10$^4$\,pps.
However, in such a case, the deuterons would not have enough kinetic energy
to exit the target or to be detected.
In this sense, the temperature, pressure and target thickness at which HeCTOr
is operated represent an ideal compromise between the two opposite
requirements.  

Future developments of the system should foresee dedicated tests to better
evaluate the rate at which the ice grows and how it depends on variables such
as the vacuum level and the gas composition.
A possible solution to overcome the problem, however, is to reheat the target
each 2-3 days during the experiment.
Such operation would allow to strongly reduce the ice thickness, though at the
expense of the LHe consumption.
For the experiment the target has been kept cold for about 11 days.
In this time interval the consumption of LHe was 2400\,L, or $\sim$\,200\,L/day.
This value could be significantly reduced by installing a system to recover
the exhaust LHe.

\subsection{Excitation energy resolution}\label{sec:energy_res}

An important observable in direct transfer reactions is represented by the
excitation energy, E*, of the heavy reaction partner.
The E* distribution is a measure of the strength of population of the
levels in the heavy residue, $^{47}$K in this case, in the transfer process.
Such strength is proportional to the cross section for each specific transfer
channel.
With the present set-up this information can be obtained from two-body
kinematics, by measuring the energy and scattering angle of the light
recoiling particles in MUGAST.
The excitation energy resolution depends on many factors (target thickness,
energy of the light ejectiles, angular and energy stragglings, intrinsic
resolution of the detectors, beam dimensions, ...) and
represents a relevant parameter for the design and analysis of an experiment.

\begin{figure}[ht]
\centering $
\begin{array}{lr}
\includegraphics[width=0.5\columnwidth]{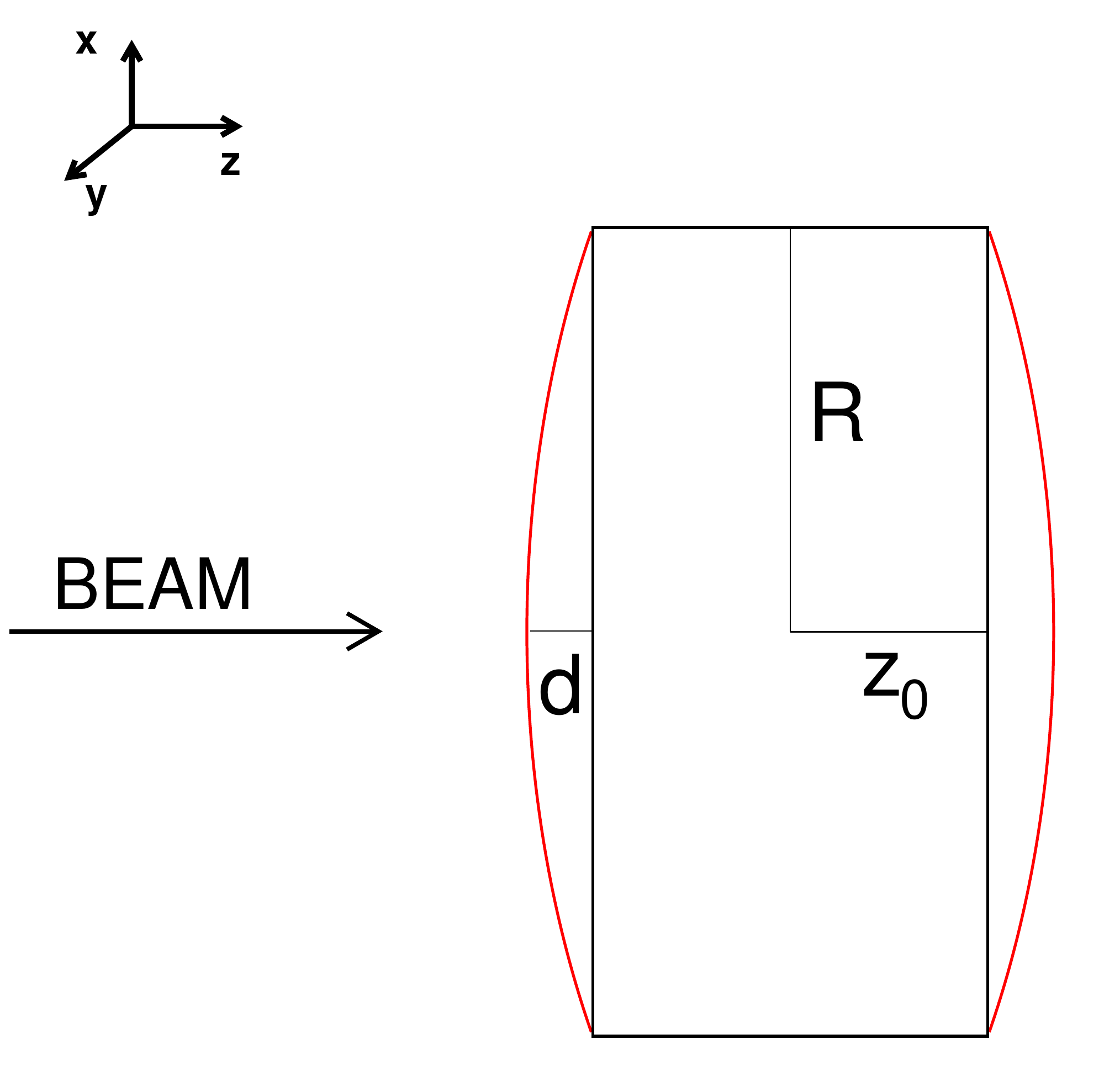} &
\includegraphics[width=0.4\columnwidth]{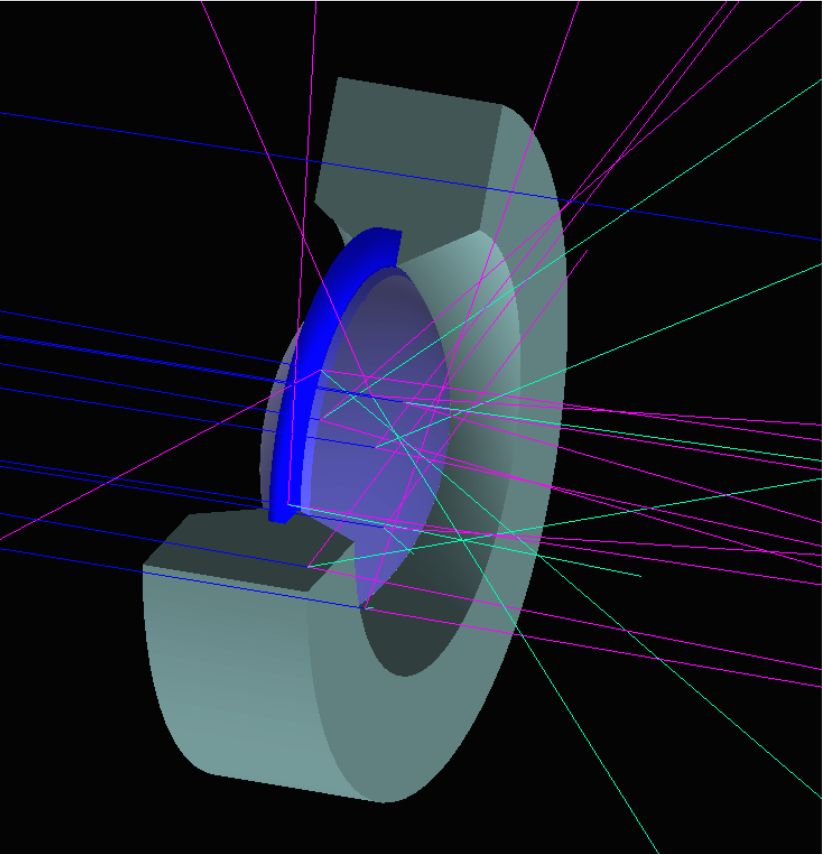}
\end{array} $
\caption{(Left) A simplified drawing of the target cell where the different
parameters of Eq.~\ref{eq:cosh} are indicated.
The red curves represent the target deformation.
(Right) An expanded view of the target generated in the Geant4 simulations.}
\label{fig:cell}
\end{figure}

The expected energy resolution for the present case has been determined
through simulations performed with \emph{nptool}~\cite{nptool}, an
open-source data analysis and Monte Carlo simulation framework developed
for low-energy nuclear physics experiments and based on the Geant4
~\cite{geant4} simulation toolkit and the ROOT~\cite{root} data analysis
framework.
In \emph{nptool} the target is created by specifying, besides the thickness,
density and type of gas, the characteristics of the front and back windows
(material, radius, thickness and possible deformations) and of the frame.
A possible parametrization for the target and window deformation is
represented by a hyperbolic cosine function.
Given the cylindrical symmetry of the target cell with respect to the beam
axis z, in the forward hemisphere (z\,$>$\,0) this function has the form:

\begin{center}
\begin{equation} \label{eq:cosh}
f(r) = z_0+(d+1)-{\mathrm{cosh}}\left[\frac{r}{R}~{\mathrm{acosh}}(d+1)\right]
\end{equation}
\end{center}

where r represents the distance from the beam axis, R is the target radius,
d the deformation at the center of the target and z$_0$ half of the target
thickness (refer to Fig.~\ref{fig:cell}, where R\,=\,8\,mm, d\,=\,0.5\,mm
and z$_0$\,=\,1.5\,mm).
In the backward hemisphere the profile is described by -$f(r)$. 

\begin{figure}[ht]
\begin{center}
\includegraphics[width=0.8\columnwidth]{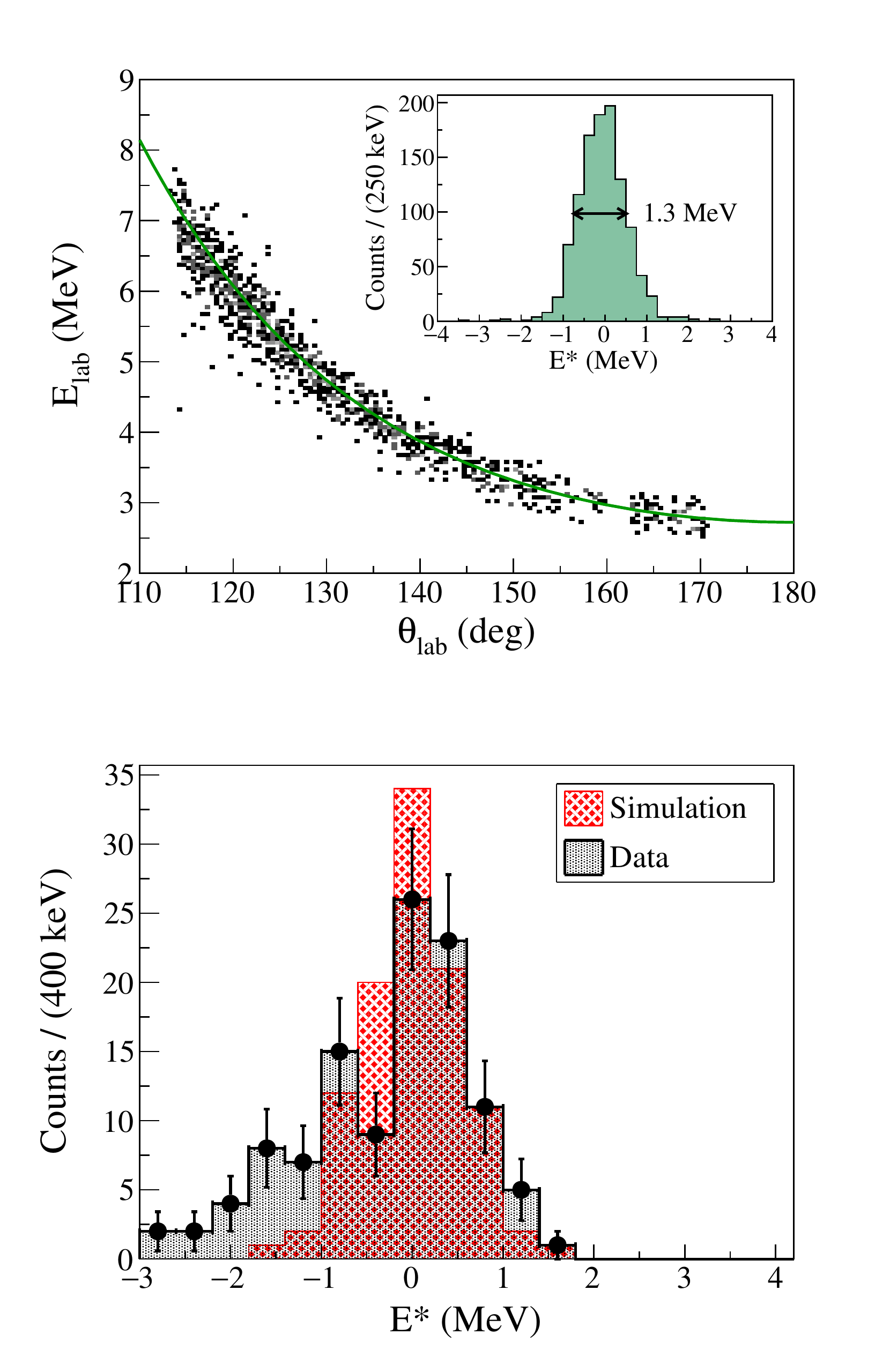}
\end{center}
\caption{(Top) Two-dimensional plot energy vs scattering angle of the deuterons
obtained in the simulation of the reaction $^{46}$Ar($^3$He,d)$^{47}$K at
E$_\mathrm{b}$\,=\,10\,MeV/u for the transfer to the ground state of $^{47}$K.
The assumed angular distribution of the emitted deuterons is isotropic in the
center-of-mass frame.
The green line represents the theoretical kinematic line.
The inset shows the resulting simulated excitation energy distribution of
$^{47}$K and the value of resolution FWHM.
(Bottom) Comparison between simulated (red) and experimental (black) E*
distributions of $^{47}$K.
The number of simulated events and the binning have been adapted for a
clearer comparison with the experimental spectrum.}
\label{fig:Ex_exp}
\end{figure}

We simulated the transfer reaction $^{46}$Ar($^3$He,d)$^{47}$K to the
ground state of $^{47}$K, in inverse kinematics at the bombarding energy
in the laboratory frame E$_\mathrm{b}$\,=\,460\,MeV (10\,MeV/u).
In a first simulation we assumed a perfectly collimated point-like beam
impinging on the target in (x,y)\,=\,(0,0).
In this way we can determine the contributions to the excitation energy
resolution coming only from the target and the detectors.
The experimental set-up here considered is composed of HeCTOr, placed 25\,mm
downstream with respect to the nominal target position, and the MUGAST array.
For the target configuration we set a thickness of 3\,mm of $^3$He gas.
In the backward hemisphere, covered by MUGAST, we set a deformation of
0.5\,mm, a 3.8-$\mu$m thick Havar window and an ice layer of 35\,$\mu$m.
The gas pressure, temperature and density are 700\,mbar, 6\,K and 4.5\,
mg/cm$^3$, respectively, corresponding to an areal density of $\sim$\,2\,
mg/cm$^2$ of $^3$He.
We assumed an isotropic angular distribution of emitted deuterons in the
center-of-mass frame.
The top panel of Fig.~\ref{fig:Ex_exp} shows the matrix where the energy of
the deuterons right after the reaction, E$_\mathrm{lab}$, reconstructed
by accounting for all the energy losses in the target and in the detector
dead layers, is plotted versus their scattering angle in the laboratory
frame, $\theta_\mathrm{lab}$.
The theoretical kinematic line is plotted in green.
The inset shows the resulting excitation energy distribution of $^{47}$K,
centered in 0 as expected for the transfer to the ground state.
The obtained excitation energy resolution is $\sim$\,1.3\,MeV FWHM.

The relative contributions to this resolution can be computed performing
different simulations considering, in each of them, only the effect of a
single target ``layer''.
We obtained
for the target thickness $\Delta \mathrm{E_t}$\,$\sim$\,0.8\,MeV,
for the target deformation $\Delta \mathrm{E_d}$\,$\sim$\,0.4\,MeV,
for the Havar window $\Delta \mathrm{E_w}$\,$\sim$\,0.6\,MeV,
for the ice layer $\Delta \mathrm{E_l}$\,$\sim$\,0.6\,MeV (all values are
FWHM).
The intrinsic resolution of the detectors is
$\Delta \mathrm{E_{Si}}$\,$\sim$\,40\,keV.
The total resolution is given by
$\Delta \mathrm{E_{tot}}=\sqrt{\Delta \mathrm{E_t}^2+\Delta \mathrm{E_d}^2+
\Delta \mathrm{E_w}^2+\Delta \mathrm{E_l}^2+\Delta \mathrm{E_{Si}}^2}$
\,$\sim$\,1.3\,MeV, in agreement with the result of the simulation where
all the effects are considered at the same time.

The experimental E* distribution can be obtained by gating on the $^{47}$K
ions detected in VAMOS and on the deuterons detected in MUGAST.
Careful considerations need to be made when comparing it with the result of
the simulation.
The experimental E* distribution is expected to contain contributions from
the different levels of $^{47}$K populated in the transfer.
The 1/2$^+$ ground state of $^{47}$K can be populated with an orbital
angular momentum transfer L\,=\,0 from $^{46}$Ar, while the first excited
states 3/2$^+$ at 360\,keV and (7/2$^-$) at 2020\,keV can be populated via
L\,=\,2 and L\,=\,3 transfer, respectively.
Assuming the spectroscopic factors for the different L transfers to be
comparable, at the most backward scattering angles in the laboratory frame,
which correspond to the forward angles in the center-of-mass frame, the
L=0 transfer to the ground state should have higher cross section than
higher L transfer and the width of the E* distribution should reflect
the experimental energy resolution.
In the bottom panel of Fig.~\ref{fig:Ex_exp} we plot in black the E*
distribution obtained with a gate only on the annular detector of MUGAST,
which covers the most backward laboratory angles
($\theta_{\mathrm{lab}}>160^{\circ}$).

To directly compare simulation and data one has to consider that the
characteristics of the beam can significantly affect shape and width of
the E* distribution.
For this reason we performed a second simulation considering more realistic
beam parameters, leaving the target characteristics unchanged from the
first simulation.
The beam profile on the xy plane is now represented by a two-dimensional
gaussian function with centroid in (x,y)\,=\,(0,0) and
$\sigma_\mathrm{x}$\,=$\sigma_\mathrm{y}$\,=\,3\,mm.
Typical spots of SPIRAL1 beams are of the order of $\sim$\,2\,mm, but
we have to consider the effect of the angular straggling of 0.9\,mrad
(computed with LISE++) occurring in the four 1.5-$\mu$m thick Mylar windows
of the CATS detector placed 2\,m before the target.
Such angular straggling has been taken into account in the simulation.
The number of events was reduced to adapt to the experimental statistics
for a more direct comparison.
The resulting E* spectrum is shown in red in the bottom panel of
Fig.~\ref{fig:Ex_exp}, superimposed to the experimental one.

The two spectra are in reasonable agreement, though the experimental
resolution is slightly larger ($\sim$\,30\,\%) than the simulated one.
This discrepancy can be explained considering that, in the experimental
data, contributions from the low-lying 3/2$^+$ state at 320\,keV in
$^{47}$K, very close in energy to the ground state, may be present.
Moreover the simulation does not consider other beam parameters which
could not be precisely determined experimentally with a single beam
profiler (CATS) detector, such as, for instance, possible impact positions
on target different from (0,0) or slight tilts in the beam direction.
The small number of counts at negative excitation energy in the experimental
distribution mainly comes from contributions from the deuteron break-up
channel after the transfer.
Due to the high cross section of the process and the limited resolution
in particle identification in MUGAST, few protons, in coincidence with
$^{47}$K in VAMOS, can be tagged as deuterons and treated as such in the
analysis.
Their energy loss in the target is then overestimated, resulting in a
negative excitation energy.

The complete experimental E* distribution and its physical implications
will be further discussed in a forthcoming publication~\cite{brugnara}.

\subsection{Transparency to $\gamma$ radiation}\label{sec:agata}

In experiments where absolute cross sections are extracted via
$\gamma$-particle coincidence, the $\gamma$-ray efficiency of the detection
set-up plays an important role in the determination of the relevant physical
quantities.
While MUGAST is nearly transparent to $\gamma$ radiation~
\cite{mugast_campaign}, the thick frame and shield of the target, if
intercepted by $\gamma$ rays before they reach AGATA, might absorb part of
them and lower the effective $\gamma$-ray efficiency.
When the nucleus produced in the reaction de-excites instantaneously through
the emission of $\gamma$ radiation,
the absorption of $\gamma$ rays by HeCTOr is minimum, since they can
pass through the entire large opening of the conic flange facing AGATA
in the backward hemisphere.
In the case that the decaying level has a long lifetime, of the order of
the nanosecond or more, the nucleus might travel a few centimeters before
de-exciting.
Part of the $\gamma$ rays will then be absorbed by the target frame before
they reach AGATA and the efficiency of the $\gamma$-ray detector will be
consequently reduced.

\begin{figure}[ht]
\begin{center}
\includegraphics[width=\columnwidth]{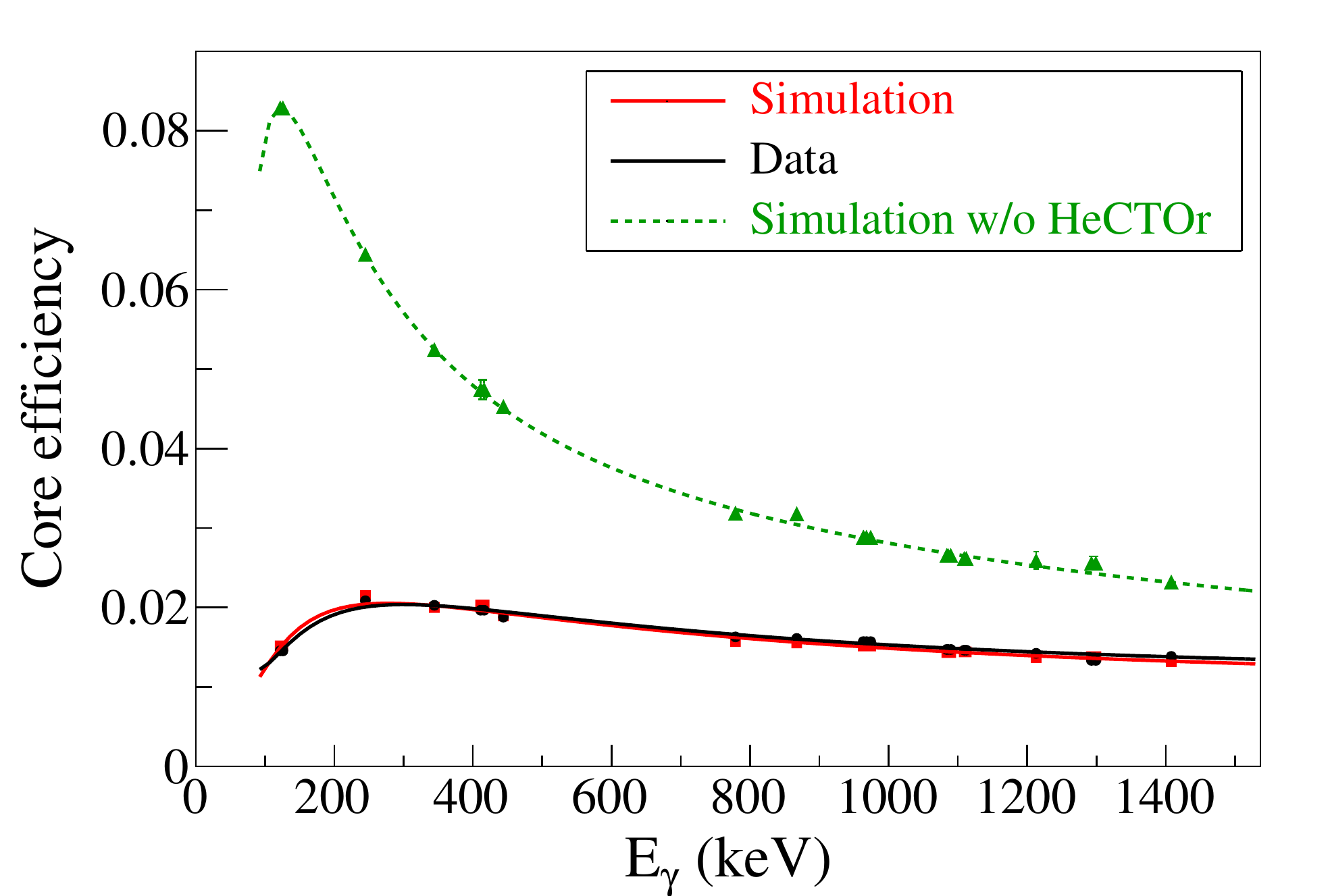}
\end{center}
\caption{Comparison of the AGATA core efficiency as a function of the
$\gamma$-ray energy between simulation (red squares) and data (black circles)
for $\gamma$ rays emitted by a $^{152}$Eu calibration source placed at 85\,mm
downstream from the target.
The green triangles and dashed line represent a second simulation where HeCTOr
is not present.
The curves are fits obtained following the prescription of Ref.~\cite{gray}.}
\label{fig:gamma_eff}
\end{figure}

Even though transparency to $\gamma$ rays was not one of the main
constraints in the design of HeCTOr, it represents an important
parameter to consider when designing and analyzing experiments
involving long-lived nuclear excited states.
In the present experiment, the 3/2$^+$  state at 320\,keV in $^{47}$K,
which decays to the ground state, has a lifetime $\tau$\,$\sim$\,1.6\,ns.
Since the average $\beta$ for $^{47}$K ions is $\sim$\,0.14, in a time
interval of the order of $\tau$ the nucleus travels approximately 6\,cm and
the $\gamma$-ray absorption due to the target can become significant.
Such effect can be quantified by computing the loss of efficiency of
AGATA when a calibration source is placed downstream with respect to the
nominal target position, thus simulating the decay of a long-lived excited
state.
Therefore we placed a $^{152}$Eu source at 85\,mm from the target and
acquired the corresponding $\gamma$-ray energy spectrum.
A simulation was then performed with the AGATA simulation code~
\cite{agata_simul}, with the $^{152}$Eu source placed in the same position,
including HeCTOr and the reaction chamber.
Figure~\ref{fig:gamma_eff} shows the comparison of the efficiency of the
cores of the AGATA crystals as a function of the $\gamma$-ray energy
between simulation (red squares) and data (black circles).
The curves are fits obtained following the prescription of Ref.~\cite{gray}.
The absolute value and trend of the efficiency as a function of E$_{\gamma}$
are in excellent agreement with the simulation.
Significantly wrong parameters for the target materials and dimensions
would in fact result not only in a different offset but also very
different shapes of the efficiency curve.
As an example, the efficiency curve obtained in a simulation where the
entire volume of HeCTOr was removed is shown in Fig.~\ref{fig:gamma_eff}
with a dashed green line.
The large difference, as compared to the other curves, clearly
reflects the effect of $\gamma$-ray absorption by the target,
which turns out to be, as expected, larger for $\gamma$ rays of lower
energies.
This result confirms the importance of carefully accounting for the effect
of $\gamma$-ray absorption by the target when analyzing the decay of
isomeric excited states.

\section{Conclusions}

In this paper we reported on HeCTOr, a thick $^3$He cryogenic target specifically
designed to be employed for direct nuclear reactions in inverse kinematics.
Among the strengths of the target are the high purity and density of scattering
centers attainable at cryogenic temperatures, where thicknesses of the order
of few mg/cm$^2$ ($\sim$\,10$^{20}$\,atoms/cm$^2$) can be obtained.
These characteristics make it particularly suited for experiments with
low-intensity radioactive beams in fragmentation and ISOL facilities.
The target has been integrated in a compact experimental set-up consisting of
three coincident detectors for a first in-beam experiment.
It showed a correct operation of its components and an excellent stability of
temperature and pressure over time.
Dedicated Geant4 simulations have demonstrated to provide good control over
different relevant experimental parameters, such as total target thickness,
energy resolution and $\gamma$-ray absorption.
Foreseen improvements of the set-up will focus on the reduction of both the
thickness of ice layers forming on the target windows and the consumption of
LHe needed to maintain the target at cryogenic temperatures.

\section*{Acknowledgments}

The authors acknowledge the GANIL staff for the invaluable support in the
phases of installation of the target and during the experiment.
We also wish to thank Emmanuel Dartois for his help with the kinetic theory
of gases.

This work was supported by the European Union's Horizon 2020 research and
innovation program under grant agreement No\,654002.
A. Gottardo acknowledges financial contribution from the MIUR PRIN 2017
call for funding, project 2017P8KMFT.

%\section*{References}

{}

\end{document}